\documentclass[conference]{IEEEtran}

\usepackage{cite}

\ifCLASSINFOpdf
 \usepackage[pdftex]{graphicx}

\usepackage{amssymb}
\usepackage{lineno}
\usepackage{lscape}
\usepackage{tabularx}
\usepackage[gen]{eurosym}
\usepackage{multirow}
\usepackage{hhline}
\usepackage{booktabs}
\usepackage{siunitx}
\usepackage{float}
\usepackage{framed}

\else

\fi

\begin{document}
%
\title{Trying to Increase the Mature Use of Agile Practices by Group Development Psychology Training\\ --- An Experiment}


\author{\IEEEauthorblockN{Lucas Gren}
\IEEEauthorblockA{Chalmers and the University of Gothenburg\\
Gothenburg, Sweden 412--92 and\\
University of S\~ao Paulo\\
S\~ao Paulo, Brazil 05508--090\\
Email: lucas.gren@cse.gu.se}
\and
\IEEEauthorblockN{Alfredo Goldman}
\IEEEauthorblockA{University of S\~ao Paulo\\
S\~ao Paulo, Brazil 05508--090\\
Email: gold@ime.usp.br}
}

\maketitle

\begin{abstract}
There has been some evidence that agility is connected to the group maturity of software development teams. This study aims at conducting group development psychology training with student teams, participating in a project course at university, and compare their group effectiveness score to their agility usage over time in a longitudinal design. Seven XP student teams were measured twice (43+40), which means 83 data points divided into two groups (an experimental group and one control group). The results showed that the agility measurement was not possible to increase by giving a 1.5-hour of group psychology lecture and discussion over a two-month period. The non-significant result was probably due to the fact that 1.5 hours of training were not enough to change the work methods of these student teams, or,  a causal relationship does not exist between the two concepts. A third option could be that the experiential setting of real teams, even at a university, has many more variables not taken into account in this experiment that affect the two concepts. We therefore have no conclusions to draw based on the expected effects. However, we believe these concepts have to be connected since agile software development is based on teamwork to a large extent, but there are probable many more confounding or mediating factors.  
\end{abstract}

\IEEEpeerreviewmaketitle


\section{Introduction}\label{sec:introduction}
Agile Project Management and its methods evolved during the nineties on ideas from lean production and more flexible product development~\cite{takeuchi1986new}, but also from practical experience saving IT projects that were about to fail \cite{sutherland2014scrum}. The main difference between lean production and agile project management is that both management ideas admit they do not know what the best end-product would look like far in advance \cite{schwaberBeedle2003agile}. The agile development processes are often intimately connected to high performing, self-managing and mature teams \cite{melnik} and the way group norms are set has been shown to increase performance \cite{teh}. Agile development, as compared to plan-driven ditto, implies more communication and focus on human factors, which make the group psychology aspects of teams a key ingredient \cite{lenbergchase}. However, the agile processes do not explicitly include the temporal perspective of what happens to all teams over time from a group maturity perspective. 

In this experiment, we conducted a longitudinal study of seven agile teams to see if the group development affects process agility. By giving half of the teams training in group psychology theory we hoped to see an effect on their measured agility. However, by only giving a 1.5-hour lecture, we did not see such an effect. We instead discuss reasons for our non-significant results and suggest next steps for future attempts at finding such effects in complex social systems.

We follow Jedlitschka, Ciolkowski and Pfahl's \cite{jedlitschka2008reporting} guidelines on how to to report experiments on software engineering throughout this paper. We will therefore start by giving a theoretical background (Section~\ref{background}), describe the experiment in detail (Section~\ref{sec:methodology}), analyze the data and show descriptive statistics and tests (Section~\ref{sec:results}), and, finally, discuss the result (Section~\ref{sec:discussion}) and provide conclusions and suggestions for future work (Section~\ref{sec:future}).

\subsection{Context}
When software development teams transition to an agile approach (i.e.\ more team-based work) more of the process is dependent on how well the team cooperates \cite{melnik}. The agile adoption sometimes fails due to the fact that an agile transition is a cultural change as well, which impose new constellations of teams \cite{iivari,tolfo2008}. To further explore the causal relationship between the group dynamics and agile practices over time, would therefore be interesting, both from a research and an industrial perspective, in order to guide agile adoptions better. 

\subsection{Problem statement}
Many aspects of group dynamics come into play in the team-based workplace \cite{wheelandev}. There are studies showing a correlation between group maturity and agile concepts (see e.g.\ \cite{Gren2015SEAA}), however, little is known of any causal relationship between them. Correlation analysis only show the connection between the two. If the mature usage of agile practices are directly dependent on group development aspects has not yet been investigated. Therefore, it would be interesting to see if group psychology training of agile software development teams could increase the adoption of concrete agile practices.

\section{Background}\label{background}

\subsection{Agile methods (processes)}
Agile methodologies can be seen as an approach rather than a technique that mostly change the culture and values behind managing projects. There are some more concrete agile methods, but they all basically share the same values. However, in order to understand how these methods work in practice, we will now shortly present some of the agile practices and how the values are implemented.    

\paragraph{eXtreme programming (XP)}\label{xp}
eXtreme programming was the first method created by the agile community and is the most researched method \cite{dybaa} and is considered relatively strict and controlled. The practices that implement the agile principles are \cite{cohen}:

\begin{enumerate}
\item \emph{The planning game.}
In the beginning of each iteration, the team, managers, and customers meet and write requirements in form of user stories (written in clear natural language and in a way that everybody can understand). During these meetings the whole group estimates and prioritizes the requirements.
\item \emph{Small releases.}
Working software is up and running and delivered very fast and new versions are released continuously, from every few days to every few weeks. 
\item \emph{Metaphor.}
Customers, managers, and developers model the system after a constructed metaphor or set of metaphors.
\item \emph{Simple design.}
Developers are asked to keep design as simple as possible. 
\item \emph{Tests.}
The development is test-driven (TDD), i.e., the test are written before the code.
\item \emph{Re-factoring.}
The code should be revised and simplified over time.
\item \emph{Pair-programming.}
All code is written by having two developers per machine. 
\item \emph{Continuous integration.}
The developers integrate new code into the system as often as possible. However, all code must pass the testing otherwise it is discarded.
\item \emph{Collective ownership.}
Developers can change code wherever necessary and the overall code is assessed.
\item \emph{On-site customer.}
A customer is in the team all the time to answer questions so the team always works according to what is needed.
\item \emph{40-hour work week.}
The team works with a sustainable pace defined as a 40 hour work week. The requirement selected for each iteration should never mean that the team needs to work overtime.
\item \emph{Open workspace.}
The team should be collocated and fit in the same room. The layout of the room should make cooperation and communication easy.
\end{enumerate}

\paragraph{Scrum}
Scrum is based on XP and is one of the more common methodologies and is built on embracing change and focus a lot on delivering value. In Scrum, the project has a prioritized backlog of requirements and use iterative development (called ``sprints'') to get basic working software for the customer to view as soon as possible. Scrum uses self-organizing teams that get coordinated through daily meetings called ``scrums.'' The manager is called a ``Scrum Master'' to clarify that it is a facilitating role and not a directive one. 

The Scrum methodology consists of three main phases: Pre-sprint planning, sprint (iteration), and post-sprint meeting. All work to be done is kept in a ``release backlog'' where from requirements (user stories) are taken to the current ``sprint backlog.'' The requirements are usually broken down from a higher abstraction level when the sprint backlog is made. The actual sprint (usually 2--4 weeks) is when the implementation is performed. Here, the sprint backlog is frozen and the team ``sprints'' to complete what was planned. The team members choose tasks they want to work on themselves. ``Scrum meetings'' also called ``Daily scrums'' are 15-minute meetings every morning were the team members check status, report problems, and keep the whole team focused on a common goal. The post-meeting is done to evaluate the process and demonstrate the current system. One important aspect of Scrum is to have small working teams in order to maximize communication, minimize overhead, and maximize the sharing of informal (or tacit) knowledge. The team should also agree and be able to define when something is considered ``done'' \cite{schwaberarticle}.

\paragraph{Lean and Kanban}
The flexible project management techniques and focus on customer value is not new. Within lean manufacturing these aspects have existed a long time (for more information about lean manufacturing see for example \cite{feld}). Many companies combine the process of Scrum with Kanban (Scrum-ban). It is important to note that Kanban is a signal card to pull products through the process within Lean production but has become a software development tool itself \cite{leansoftware}. Scrum is a more strict process and can be modified by changing the WIP (work in progress) in each sprint into being connected to the work-flow state to prevent too much WIP. Kanban also allows adding items within each sprint. Another aspect is to change the sprint backlog owned by the team into a Kanban board with multiple teams with work-flow state instead. The Kanban board is never reset after a sprint and can be followed over time, and is also less dependent on collocation. Scrum only allows three different roles of the team, while Kanban does not have a limit. Therefore, larger teams in larger organization with a diversity of specializations often use Kanban or Scrum-ban when possible \cite{scrumban}.

\paragraph{Crystal}
We will not describe the Crystal methodologies in detail but, generally speaking, they are built on the assumption that the main problem in software development is poor communication. Crystal focuses on people, interaction, community, skills, talents, and communication as main effects on performance \cite{cockburn}.

The twelve agile principles are a very high-level description of a work environment. Agile software development is an ambiguous concept with descriptions on various levels of abstraction. Many of these are obviously connected to group dynamics. The problem is that these psychological aspects are not described in detail in the methods (processes). This means that this dimension is left out for practitioners to figure our for themselves to a large extent. In order to try to operationalize agility and correlate the measurement to group maturity over time, we enforced the twelve original XP practices (described in Section~\ref{xp}) on all the participating student teams and then opted to use the Perceptive Agile Measurement developed by So and Scholl \cite{so} in order to measure this ``agile'' behavior over time. All the items are included in Section~\ref{Items}.

\subsection{Groups and Teams}
A group can be defined as: ``three or more members that interact with each other to perform a number of tasks and achieve a set of common goals'' \cite{grupp}. If the group is larger all the members might not have a common goal, which means that larger groups often consist of subgroups. Some studies have shown that smaller groups are more productive than larger groups with a threshold at around eight individuals~\cite{wheelan2009}. In psychology, a ``work-group'' is a group that has a shared view of the group goal and has developed a structure that enables goal achievement. A team, on the other hand, is an effective work-group, however, we will use the terms somewhat interchangeably in this paper, since agile work-groups are called ``teams'' no matter their actual effectiveness. In social psychology, though, only 17\% of all groups were considered teams according to one study~\cite{wheelan}.

The group research in psychology received much attention after the second world war and before the sixties. After that, the focus in research was on the individual instead of groups \cite{wheelan}. The start of the human factors research in software engineering has also mostly focused on individuals and their personalities and traits for 40 years without finding any coherent results~\cite{forty}. Therefore, we have reason to believe that much of what happens in software engineering is set on team-level, which means that ``agility'' is hard to obtain if we do not understand the group dynamics of agile teams, or as Wheelan and Hochberger~\cite{wheelan} very adequately put it: ``before one jumps to fix something, one has to know what is broken.''

During so many years of research on groups in psychology, there are, of course, a diversity of group development models \cite{wheelan1993}. However, there seems to be a reoccurring patterns of what happens to all types of groups when humans get together in order to solve a task. The first researchers to aggregate models into a general group development model were Tuckman and Jensen~\cite{tuckman} in the seventies. In the nineties Susan Wheelan did a similar aggregation of existing models that resulted in the Integrated Model of Group Development that we used in this study. However, Tuckman and Jensen's~\cite{tuckman} model with the phases; Forming, Storming, Norming, and Performing correspond well to the stages suggest by Wheelan \cite{wheelan}.

\subsection{Wheelan's Integrated Model of Group Development}\label{sub:integratedgroup}
\begin{figure*}
\centerline{\includegraphics[width=130mm]{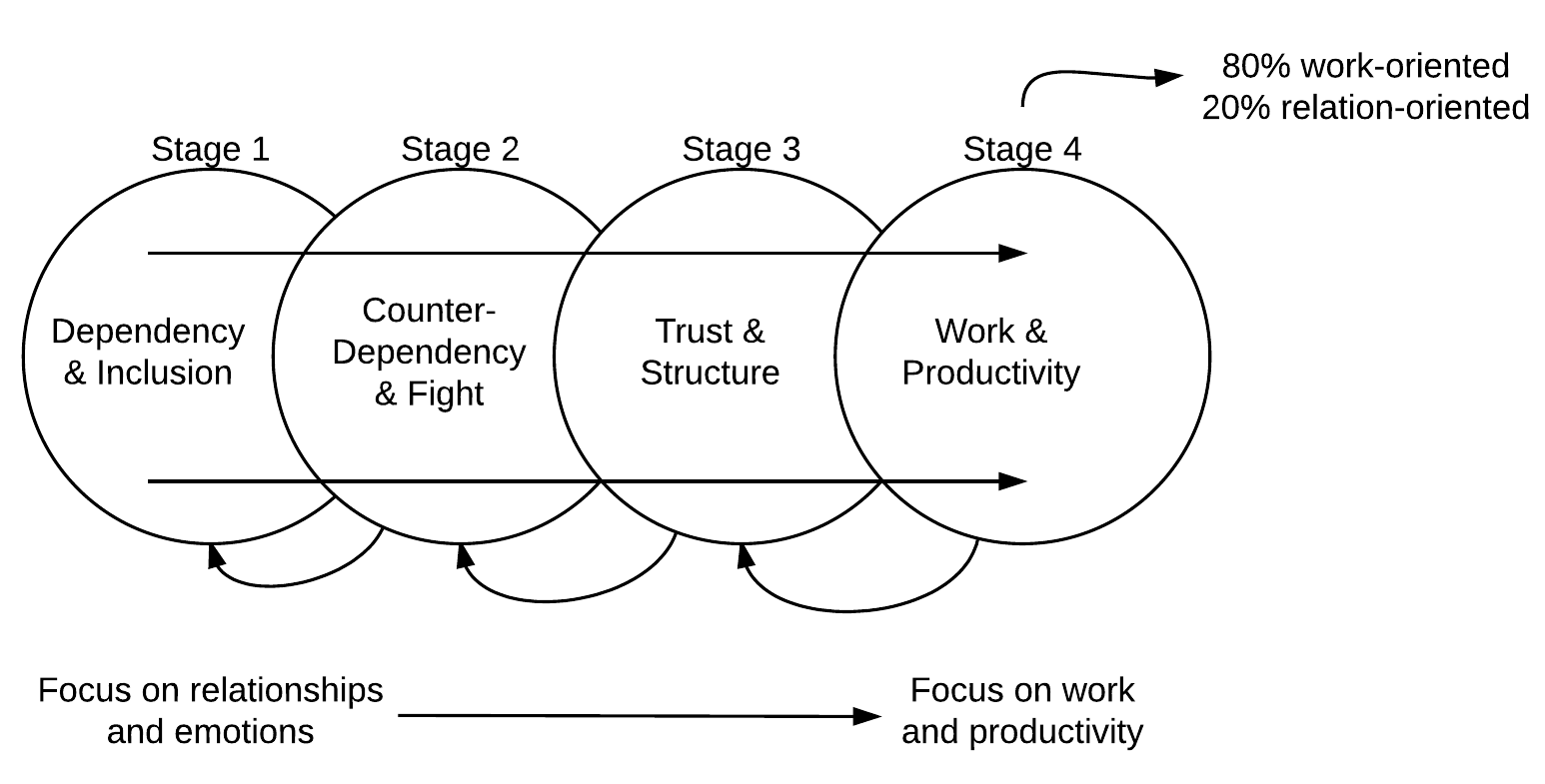}}
\caption{The Group Development Stages~\cite{wheelan2012}}
\label{fig:groupstages}
\end{figure*}

The Integrated Model of Group Development (or IMGD) describes four different stages that all groups go through in their journey towards becoming a well-functioning high performing team. These stages are illustrated in Figure~\ref{fig:groupstages} and described next. The Group Development Questionnaire (the GDQ), that is a measurement of how much energy the group is spending on each development stage, is described afterward.

\paragraph{Stage 1 --- Dependency and Inclusion}
During the first stage of group development (i.e.\ when the group is new) the group members have more focus on safety and inclusion, a dependency on the designated leader, and more of a wish for order and structure, than in more mature stages. A group at stage one can still get work done, but will focus more on figuring out who the other people are. There is a lack of structure and the group needs to become organized, being able to do efficient work, and achieve the group goals. The group members need to create a sense of belonging and lay the foundation for how to interact within the group. At this first stage, there is a lack of the feeling of belonging to a group, but after this stage people start feeling safe enough to state their ideas and contribute to how they think the group should work in order to achieve its goals. If this does not happen groups stagnate, which is often noticed when group members stop doing work between meetings and even stop attending the group meetings~\cite{wheelan}.

\paragraph{Stage 2 --- Counter-Dependency and Fight}
During the second stage the group starts having conflict. These differences in opinion is a must in order to create clear roles based on real competence and to make it possible to work together in a constructive manner. The group members have to go through this more turbulent stage in order to build trust. After feeling safe and therefore daring to have these conflicts, a sense of loyalty emerges, which is needed to create cohesion. Since we do not have a clear picture of goals and roles in the beginning, we need this emotional and hard work in order to get shared perceptions of values, norms, and goals, which need to be set on group-level. Since everybody needs to believe in the group values and norms for them to fill their purpose, the rules of the game need to be negotiated so that all members thoroughly believe in them. The more shallow discussions about goals probably present in the first stage, will now be more emotional or include disagreements~\cite{wheelan}. 

\paragraph{Stage 3 --- Trust and Structure}
During the third stage the structure is getting into place and the roles are now actually based on competence instead of status, power, or safety concerns. The communication patterns are more open and also more task-oriented. In this stage the role, organization, and process negotiations are most often more mature and there will be an evident clarification and consensus regarding the group goals. The group members also spend time solidifying positive relationships, and when the tasks are adjusted to competence the likelihood of goal achievement is higher. At this stage the leader's roll goes from needing to have been more directive to being more consultative. The communication structure is also more flexible (i.e.\ group members talk to whomever they need). Along the group development the content of the communication is also more and more task-oriented instead of relation-oriented. Groups always need the relation-oriented communication since we always need to do the maintenance of discussing how we work together as a group. Therefore, conflict will still occur but be over much faster since the group has better conflict management techniques. Work satisfaction and cooperation increase together with cohesion and trust. At this stage the individual commitment to the group goal will be higher (i.e.\ we care about what the group is doing on a personal level). This means we will see a voluntary conformity with the norms and helpful deviation from the group (like sub-grouping) will be accepted if considered helpful for the group as a whole~\cite{wheelan2012}.

\paragraph{Stage 4 --- Work and Productivity}
The forth stage of group development is when the group does even better with regards to the purpose of stage three. This means that the group focuses on getting the task done well together as well as maintaining group cohesion over a longer period of time. It is important to realize that there is a large set of variables that can and will disturb the group development. Basically, all changes will have such an effect, e.g.\ change of demands from the organization, losing staff, getting new staff, and so on and so forth. The challenge in stage four is therefore to try to maintain the effectiveness reached, and the most effective groups do not discuss task-related issues a hundred percent of the time, but actually, still spend around twenty percent discussing how to work together, which is key when maintaining high performance. The characteristics of the decision-making in such teams are participatory and the team encourages task-related conflicts, since they help finding better solutions to problems faced. A person who is or has been on such a team will recognize the intensity of the work and the effectiveness together with a very high interpersonal attraction between group members. People in such high performing teams often look at their work with excitement and joy and getting work done is easy and members have the feeling of being a part of the absolute best team in the world. Getting to stage four takes a lot of work both from internal group members but the group also needs to be given the right conditions from their surrounding ecosystem~\cite{wheelan2012}.

\subsection{The Group Development Questionnaire (GDQ)}\label{group}
Wheelan~\cite{wheelan2012} was not the first one who found these characteristic stages of group development, but she contributed with a tool to measure these different stages with four scales put together in a questionnaire. Her tool has made it possible to measure and therefore diagnose where a specific group is focusing its energy from a group developmental perspective. The survey has a total of 60 items and provides a powerful tool for research and interventions in teams. Scale four (GDQ4) is the ``work and productivity'' and has been shown to correlate with a set of effectiveness measures in different fields. Examples are that groups that have high scores on GDQ4 finish projects faster~\cite{wheelan1998}, students perform better on standardized test (SAT scores) if the faculty team scores high on GDQ4~\cite{wheelan1999, wheelan2005}, and intensive care staff save more lives in surgery \cite{wheelan20032}.

\subsection{Technology under investigation}
The group development stages have been well-known for many years in social psychology \cite{wheelandev}. Helping teams to develop and mature in their group development have been shown to increase productivity and effectiveness in a diversity of fields (see Section~\ref{group}). Therefore, we want to see if helping teams to mature from a group psychology perspective also gets them to mature in their usage of the XP practices. 

There are many group development models, but few have been scientifically validated in the way the group development questionnaire (GDQ) has \cite{wheelan}. Since correlations have been found between the group score on the ``work and productivity'' scale of the GDQ and other external effectiveness measurements, it would interesting to explore its effect on agile software development teams in their adoption of agile practices. Especially since agile methods have been shown to increase software development project success \cite{serrador2015does}. 

Evidence-based interventions within group development with the GDQ have been shown to increase the group maturity in teacher teams \cite{jacobsson2009increasing}, and to increase the velocity of group development \cite{jacobsson2010group}. 

\subsection{Relevance to practice}
If group development training can be shown to increase the agility of software engineering teams, such aspects would be appropriate to explicitly integrate into the implementation of ``agility'' in all organizations conducting software development.

\section{Experiment planning}\label{sec:methodology}

\subsection{Goals}
The goal of this experiment was to see if training an agile team in group developmental psychology would increase their agility through more mature use of the agile practices. 

\subsection{Experimental units}
In order to conduct an experiment the study was conducted with 43 students in an agile software development course at the University of  S\~ao Paulo. Group developmental aspects apply to all group-work and therefore working with students as research subjects is a valid representation of software development conducted by developers on all knowledge levels. However, we would still be careful in generalizing a result to a larger population than that of developers in the phase of learning an agile approach (i.e.\ individuals with little experience of agile software development in practice). The course were offered to 3rd year students, however, most students usually take the course in the 4th or 5th (last) year of their software engineering degree. The course is also open to graduate students who are given the possibility to take the course twice during their graduate education.

The student teams in this study comprised students enrolled in a project XP software development course called ``The Laboratory of XP'' at the Institute of Mathematics and Statistics at the University of S\~ao Paulo. The purpose of the course is to introduce agile methods through the use of XP. These methods included, at a minimum, the twelve practices presented in Section~\ref{xp}. Some other staff at the university acted as customers and had to pitch their project ideas to the students, who signed up for the most interesting one from their point of view. All the teams included six to eight members and a more experienced student acting as a an agile coach for the team. The process was put together by the student teams themselves and we allowed any type of additional practices they selected as long as it was within the XP framework. As an example, we enforced collocation of a minimum of eight hours per week.

\subsection{Experimental material}
The experimental object was the agile software development team. Group norms and cooperation are set on group level and therefore the actual ``team'' is the relevant level of analysis. 

\subsection{Tasks}
The experimental tasks applied in this experiment was for the teams (one team at a time) to listen and reflect on group development theory and discuss its applicability in connection to their own team.  

\subsection{Hypotheses, parameters, and variables}\label{Items}
The construct used to measure agile practices and the behavior connected to them, was the mature usage of nine agile practices as defined by So and Scholl \cite{so}:

\begin{framed}
\scriptsize \textbf{Iterative Planning:} (1) All members of the technical team actively participated during iteration planning meetings. (2) All technical team members took part in defining the effort estimates for requirements of the current iteration. (3) When effort estimates differed, the technical team members discussed their underlying assumption. (4) All concerns from team members about reaching the iteration goals were considered. (5) The effort estimates for the iteration scope items were modified only by the technical team members. (6) Each developer signed up for tasks on a completely voluntary basis. (7) The customer picked the priority of the requirements in the iteration plan.
\end{framed}
   
\begin{framed} 
\scriptsize \textbf{Iterative Development:} (1) We implemented our code in short iterations. (2) The team rather reduced the scope than delayed the deadline. (3) When the scope could not be implemented due to constraints, the team held active discussions on re-prioritization with the customer on what to finish within the iteration. (4) We kept the iteration deadlines. (5) At the end of an iteration, we delivered a potentially shippable product. (6) The software delivered at iteration end always met quality requirements of production code. (7) Working software was the primary measure for project progress.
\end{framed}
    
\begin{framed}
\scriptsize \textbf{Continuous Integration and Testing:} (1) The team integrated continuously. (2) Developers had the most recent version of code available. (3) Code was checked in quickly to avoid code synchronization/integration hassles... (4) The implemented code was written to pass the test case. (5) New code was written with unit tests covering its main functionality. (6) Automated unit tests sufficiently covered all critical parts of the production code. (7) For detecting bugs, test reports from automated unit tests were systematically used to capture the bugs. (8) All unit tests were run and passed when a task was finished and before checking in and integrating. (9) There were enough unit tests and automated system tests to allow developers to safely change any code.
\end{framed}
 
\begin{framed}
\scriptsize \textbf{Stand-Up Meetings:} (1) Stand up meetings were extremely short (max. 15 minutes). (2) Stand up meetings were to the point, focusing only on what had been done and needed to be done on that day. (3) All relevant technical issues or organizational impediments came up in the stand up meetings. (4) Stand up meetings provided the quickest way to notify other team members about problems. (5) When people reported problems in the stand up meetings, team members offered to help instantly.
\end{framed}

\begin{framed}
\scriptsize \textbf{Customer Access:} (1) The customer was reachable. (2) The developers could contact the customer directly or through a customer contact person without any bureaucratic hurdles. (3) The developers had responses from the customer in a timely manner. (4) The feedback from the customer was clear and clarified his requirements or open issues to the developers.
\end{framed}

\begin{framed}
\scriptsize \textbf{Customer Acceptance Tests:} (1) How often did you apply customer acceptance tests? (2) A requirement was not regarded as finished until its acceptance tests (with the customer) had passed. (3) Customer acceptance tests were used as the ultimate way to verify system functionality and customer requirements. (4) The customer provided a comprehensive set of test criteria for customer acceptance. (5) The customer focused primarily on customer acceptance tests to determine what had been accomplished at the end of an iteration.
\end{framed}
    
\begin{framed}
\scriptsize \textbf{Retrospectives:} (1) How often did you apply retrospectives? (2) All team members actively participated in gathering lessons learned in the retrospectives. (3) The retrospectives helped us become aware of what we did well in the past iteration/s. (4) The retrospectives helped us become aware of what we should improve in the upcoming iteration/s. (5) In the retrospectives (or shortly afterwards), we systematically assigned all important points for improvement to responsible individuals. (6) Our team followed up intensively on the progress of each improvement point elaborated in a retrospective. 
\end{framed}

\begin{framed}
\scriptsize \textbf{Collocation:} (1) Developers were located majorly in... (2) All members of the technical team (including QA engineers, db admins) were located in... (3) Requirements engineers were located with developers in... (4) The project/release manager worked with the developers in... (5) The customer was located with the developers in...
\end{framed}

The group maturity (or effectiveness) operationalization was done through using Scale 4 of the GDQ \cite{wheelan}. All the items in the GDQ scale cannot be shared here due to copyright, however, we can include three example items:

\begin{itemize}
\item The group gets, gives, and uses feedback about its effectiveness and productivity.
\item The group acts on its decisions.
\item This group encourages high performance and quality work.
\end{itemize}

The group development measurement on Scale 4 was assessed on a 5-point Likert scale (1 = low agreement to the statement and 5 = high agreement). The agile items were assesses on a 7-point Likert scale (1 = never and 7 = always). These scales were used for the simple reason that these measurements were developed and validated using these exact scales. 

Both measurements have been validated using a factor analysis \cite{fabrigar} and a test for internal consistency (using the Cronbach's $\alpha$ \cite{cronbach}).

The formal research hypothesis for each scale is that the mean values for the scale is different between the two measurements, or $H_{1}: \mu _{1} \neq \mu _{2}$.

\subsection{Design}
We used a longitudinal research design in order to test differences in group mean value scores on the two measurements over time. The first measurement comprised seven teams and 43 student responses, and the second measurement comprised the same seven teams with 40 responses, i.e.\ three student were absent during the second measurement.

\subsection{Procedure}
The two measurement surveys were distributed to the teams five weeks into their software development projects (during their scheduled and collocated development sessions). The reason was to let the students actually form teams and have done some work before the first measurement. Three of the participating seven teams were randomized into the experimental group and the remaining four teams were used as a control group. The randomization was done by first writing the numbers ``3'' and ``4'' on paper slips and letting a person not connected to the experiment draw one folded slip for the research group (three groups were selected). The second step was conducted by writing all team names on other paper slips and letting the person draw three slips to be used for the research group. 

On week six, the three selected teams participated in a 1.5-hour group development training with a discussion on the applicability to their own team. During the first hour of the training, the first author of this paper presented The Integrated Model of Group Development \cite{wheelandev} and its four group developmental stages. The idea is, briefly, that there are predictable group developmental stages that all groups have to go through in order to work effectively. If team-members are aware of these there is a smaller probability of the team getting stuck on group issues, which leads to quicker and higher quality work \cite{wheelandev}. Aspects covered were, for example, goal-setting, role clarification, decision-making, and leadership issues of groups in different development stages. 

On week eleven, the second measurement was conducted using the same procedure as in the first measurement. 

\subsection{Analysis procedure}
The data was analyzed using a general linear model for repeated measures (i.e.\ a standard repeated measures ANOVA). Such a model assumes normality in data, but since we did not find any significant result, we did not proceed to use non-parametric tests (since these are more restrictive and would therefore neither show any significance).




\section{Results}\label{sec:results}

\subsection{Descriptive statistics}
Since we aimed at affecting the agile practices score by conducting group psychology training, we first looked at if we managed to increase the group dynamics score. Since that was not the case we already knew we did not succeed with the intended plan of the experiment. However, we still looked for differences in the agile practices measurement to see if they differed anyways between the two measurements. The only two significant differences we found between the first and second measurements were that the scales ``Retrospectives'' and ``Customer Acceptance Tests.'' Therefore, we show the descriptive statistics for these scales as well (see Table~\ref{fig:desc}).

\begin{table*}
\renewcommand{\arraystretch}{0.9}
\caption{Descriptive statistics.}
\label{fig:desc}
\centering
\begin{tabular}{c||c||c||c||c||c||c||c}
\hline
\bfseries   & Research Group & Mean & Std.\ Deviation & N\\
\hline\hline
GDQ4 1st Measurement & Yes & 3.9226 & .22378 & 3\\
\hline
GDQ4 2nd Measurement   & Yes & 3.8579 & .92728 & 3\\
\hline
GDQ4 1st Measurement & No & 3.9007 & .19832 & 4 \\
\hline
GDQ4 2nd Measurement   & No & 4.0055 & .33619 & 4\\
\hline
Retrospectives 1st Measurement  & Yes & 3.4963 & 1.54921 & 3\\
\hline
Retrospectives 2nd Measurement & Yes &  5.8500 & .42517 & 3\\
\hline
Retrospectives 1st Measurement  & No & 4.9280 & 1.05903 & 4\\
\hline
Retrospectives 2nd Measurement & No & 5.9099 & .54201 & 4\\
\hline
Cust.\ Accept.\ Tests 1st Measurement  & Yes & 3.6319 &  .38193 & 3\\
\hline
Cust.\ Accept.\ Tests 2nd Measurement  & Yes & 4.6400 & .90598 & 3 \\
\hline
Cust.\ Accept.\ Tests 1st Measurement  & No  & 4.3349 & .93600 & 4 \\
\hline
Cust.\ Accept.\ Tests 2nd Measurement  & No  & 4.8229 & .91841 & 4\\
\hline
\end{tabular}
\end{table*}

\subsection{Data set preparation}
A mean value was calculated based on the collected data for each individual, and then for the team according the agile practices as defined by So and Scholl \cite{so}. The measured agile practices were: Iteration planning, Iterative development, Continuous integration and testing, Stand-up meetings, Customer access, Customer acceptance tests, Retrospectives and Collocation. The group development Scale 4 individual items were also turned into a mean value for each individual and then for the groups separately. Since we wanted to run the analysis on group-level we only have three mean values in the research group and four values in the control group (seven groups in total).

\subsection{Hypothesis testing}
Since we have so few data points, we cannot assess the population distribution based on our sample. However, other studies have shown this kind of data to be normally distributed \cite{wheelan,so}. Also, since we did not find any significant results based on parametric tests, neither would we for non-parametric tests (since they are more restrictive). We began by testing the group effectiveness score (GDQ4 mean values) for the first and second measurements and can conclude that we did not see a significant effect (see Table~\ref{fig:gdq_test}). 

\begin{table*}
\caption{ANOVA for the two repeated GDQ4 measurements.}
\centerline{\includegraphics[width=120mm]{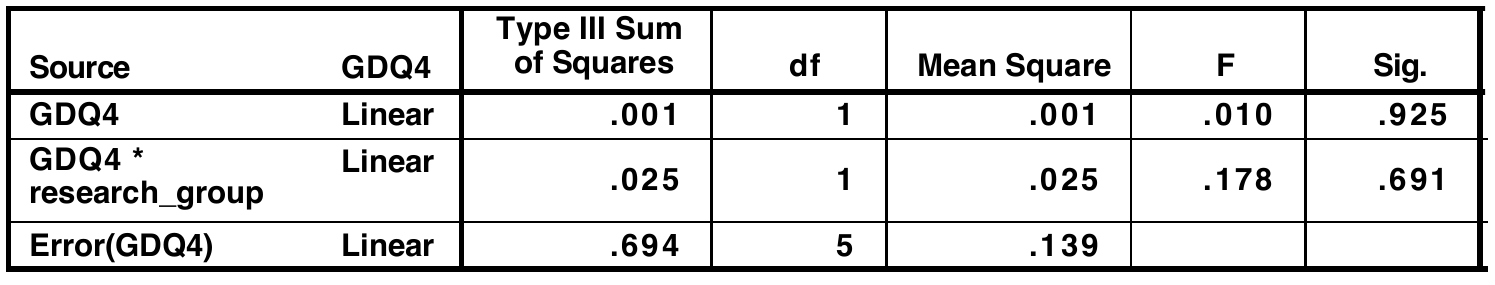}}
\label{fig:gdq_test}
\end{table*}
We then, still, ran the same analysis for all the agile practices and only found that the scales ``Retrospectives'' and ``Customer acceptance tests'' were different between the two measurements overall and not in connection to whether they were in the research group or not (see Table~\ref{fig:r_test} and Table~\ref{fig:cat_test}).

\begin{table*}
\caption{ANOVA for the two repeated Retrospectives measurements.}
\centerline{\includegraphics[width=120mm]{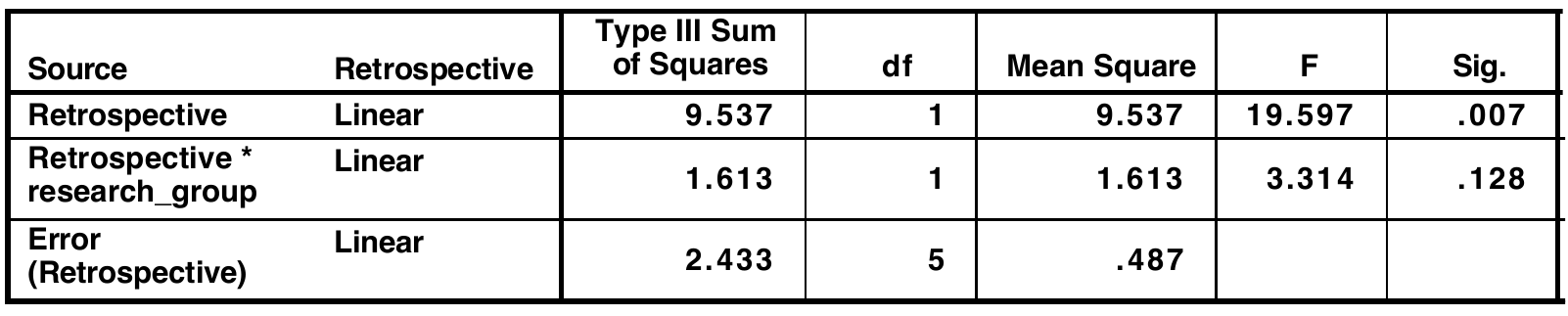}}
\label{fig:r_test}
\end{table*}

\begin{table*}
\caption{ANOVA for the two repeated Customer acceptance tests measurements.}
\centerline{\includegraphics[width=120mm]{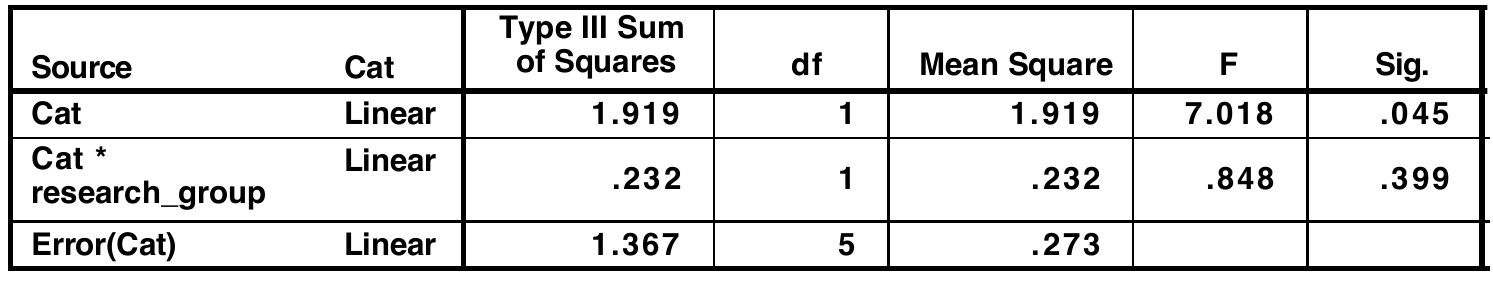}}
\label{fig:cat_test}
\end{table*}

We conclude that the group development effectiveness measurement (GDQ Scale 4) was not different between the research group and the control group (not in the first nor the second measurement). The two agile practices ``Retrospectives'' and ``Customer acceptance tests'' where both different overall between the two measurements, but not depending on if the teams were in the research group or the control group.

\section{Discussion}\label{sec:discussion}

We did not find any of the expected results in this study. Clearly, just having 1.5 hours of training and discussion is not enough to help the group to develop, even if 1.5 hours of a workweek of 8 hours (like the students in the course had) would be equivalent to 7.5 hours of working full-time 40 hours a week. When taking a closer look at when other experiments succeeded in significantly helping the groups to develop and capture their increased maturity, it turns out they did a much larger intervention then was applied in this study.  Jacobsson and Wramsten Wilmar \cite{jacobsson2009increasing}, for example, gave the groups eight different interventions of group development assessment and enforced improvement points that the group had to work on until the next workshop, plus the students were fully dedicated to only one course. In hindsight, we probably would have needed something similar in order to move the groups forward in the research group. There is also, of course, the possibility of software engineering teams' agility not being as dependent on group maturity as we might think.

\subsection{Threats to validity}
We believe the layout of the experiment has potential. Of course, even if we would have found a significant difference, we would still have to have been careful when generalizing the results due to the very small sample size. Regarding construct validity the first author was present during the data collection and could answer any potential questions regarding the questionnaire. However, since we did not want the control group to get training of group development we provided all participants with as little information as possible before the first survey, since we did not want to introduce bias. The trade-off is of course that the participants could have misinterpreted the questions and failed to answer in connection to our intended operationalization of constructs. Regarding learning effects between measurement, the GDQ has been shown to be stable for repeated measurements as such \cite{wheelan}. We have no such studies for the agile practices, which means that we might have seen a learning effect when students answer that part of the survey.

In order to prevent hypothesis guessing, we only informed the participant that the research was about looking at connections between group psychology and agile practices and not more detail on how we expected them to be connected. The internal validity is considered quite high in this experiment since we used validated scales as defined and validated quantitatively by other researchers \cite{wheelan,so}. However, inter-group communication between the research groups and the control groups is also a threat our experimental research design.  

We draw no inference from this experiment. We do not want to state that group development causes more mature use of agile practices, nor the opposite. 

\subsection{Lessons learned}
The largest lesson learned from this experiment is evidently to check the level of intervention effort needed to move groups forward in their development before conducting this kind of an experiment. We still do not known the effort needed, but the span is more then one 1.5 hours workshop with a second measurement two months later, and less than six to eight workshops of 2--3 hours during a full year with connected action plans and follow-up. By having more time with the teams we could have focused even more concretely on, for example, goal-setting, role clarification, decision-making, functional sub-grouping, or leadership issues, like in \cite{jacobsson2009increasing}. 

\section{Conclusions and future work}\label{sec:future}
We obtained an insignificant result of this experiment. We therefore have no conclusions to draw based on the expected effects. However, we believe these concepts could still be connected since agile software development is based on teamwork to a large extent. We evidently need a larger intervention effort and, of course there could also be more confounding or mediating factors we have not thought of in the context of agile software development teams. 

We would like to redo this experiment with more resources and be able to give the teams in the research group eight times more workshops with connected action plans in order to see if we can get a similar effect as has been shown with teacher teams \cite{jacobsson2009increasing}. It would, of course, be advantageous to include as many teams as possible and at multiple universities and companies to increase the statistical power of the experiment. 

\bibliographystyle{IEEEtran}

\bibliography{references}

\end{document}